\documentclass[aps,prl,reprint,groupedaddress]{revtex4-1}
\usepackage{graphicx} 
\usepackage{dcolumn} 
\usepackage{bm}
\usepackage{hyperref}

\begin{document}

\title{Unexpected neutron/proton ratio and isospin effect in low-energy antiproton induced reactions}
\author{Zhao-Qing Feng}
\email{Corresponding author: fengzhq@impcas.ac.cn}

\affiliation{Institute of Modern Physics, Chinese Academy of Sciences, Lanzhou 730000, People's Republic of China}

\date{\today}

\begin{abstract}
The inclusive spectra of preequilibrium nucleons produced in low-energy antiproton-nucleus collisions are thoroughly investigated within the framework of the Lanzhou quantum molecular dynamics model for the first time. All possible reaction channels such as elastic scattering, annihilation, charge exchange and inelastic scattering in antibaryon-baryon, baryon-baryon and meson-baryon collisions have been included in the transport model. The unexpected neutron/proton ratio in comparison to the pion and proton induced reactions is caused from the isospin effects of pion-nucleon collisions and the density dependence of symmetry energy. It is found that the $\pi^{-}$-neutron collisions enhance the neutron emission in the antiproton annihilation in a nucleus. Different to heavy-ion collisions, the isospin effects are pronounced at the low kinetic energies. A soft symmetry energy with the stiffness of $\gamma_{s}=$0.5 at subsaturation densities is constrained from the available data of the neutron/proton spectra.

\begin{description}
\item[PACS number(s)]
25.43.+t, 24.10.-i, 21.65.Ef
\end{description}
\end{abstract}

\maketitle

Since the first evidence of antiprotons was found in 1955 at Berkeley in collisions of protons on copper at the energy of 6.2 GeV by Chamberlain, Segr\`{e}, Wiegand and Ypsilantis \cite{Ch55}, the secondary beams of antiprotons were produced at many laboratories, such as CERN, BNL, KEK etc \cite{Ch57,Ag60,Le80,Ea99}. The stochastic cooling method provides the possibility for storing the antiprotons produced in proton-nucleus collisions. The particles $W^{\pm}$ and $Z^{0}$ were found for the first time with the high energy protons colliding the stored antiprotons at CERN \cite{Ua83}. On the other hand, the antiproton-nucleus collisions are motivated to many interesting issues, i.e., charmonium physics, strangeness physics, antiprotonic atom, symmetry, in-medium properties of hadrons, cold quark-gluon plasma, highly excited nucleus etc \cite{Am02,Ra80}. Recently, the antiproton-antiproton interaction was investigated by the STAR collaboration in relativistic heavy-ion collisions \cite{Star}.

In the past decades, the nuclear reactions induced by antiprotons were investigated with the facilities of the low-energy antiproton ring (LEAR) at CERN \cite{Go96}, the National Laboratory for High Energy Physics at KEK \cite{Mi84} and Brookhaven National Laboratory Alternating Gradient Synchrotron (AGS) accelerator \cite{Le99}. A number of interesting phenomena were observed, e.g., the delayed fission from the decay of hypernuclei in antiproton annihilations on heavy nuclei \cite{Bo86}, unexpected enhancement of the $\Lambda/K^{0}_{S}$ ratio \cite{Mi88}, decay mode of highly excited nucleus etc \cite{Eg90,Ja93}. The low-energy antiprotons usually annihilate at the nucleus surface because of the large absorption cross section. The huge annihilation energy are available for producing the 2-6 pions \cite{Ho94,Lu02}. The subsequent processes are complicated and also associated with the multiple pion-nucleon interaction, which result in the fragmentation of target nucleus and the preequilibrium emissions of complex particles \cite{Ki96,Lo01}.

The dynamics of the antiproton-nucleus collisions is more complicated in comparison to hadron (proton, $\pi$, $K$ etc) induced reactions and to heavy-ion collisions, in which the particles produced in the annihilation of the antiproton in a nucleus are coupled to the subsequent collisions with surrounding nucleons. The dynamics of antiproton-nucleus collisions is complicated, which is associated with the mean-field potentials of hadrons in nuclear medium, and also coupled to a number of reaction channels, i.e., the annihilation channels, charge-exchange reaction, elastic and inelastic collisions. There has been several approaches for describing the nuclear dynamics induced by antiprotons, e.g., the intranuclear cascade model \cite{Cu89}, kinetic approach \cite{Ko87}, Giessen Boltzmann-Uehling-Uhlenbeck transport model \cite{La09}, Statistical Multifragmentation Model \cite{Bo95} and the Lanzhou quantum molecular dynamics (LQMD) approach \cite{Fe14}. Part of experimental data can be understood within these models.

A more localized energy is deposited in the nucleus with an excitation energy of several hundreds of MeV. The hot nucleus proceeds to the explosive decay via multifragmentation process or the sequential particle evaporation. On the other hand, the collisions of the antiproton and secondary particles with surrounding nucleons lead to the pre-equilibrium particle emissions, which are related to the scattering cross sections of each reaction channels, antiproton-nucleon interaction, particle-nucleon potentials, density profile of target nucleus. The unexpected large neutron yields produced by stopped antiprotons in nuclei were reported in the LEAR experiments \cite{Po93,Po95}. The phenomena has been puzzling physicists several decades \cite{Ps95}.

In this letter, the preequilibrium nucleon emissions and the neutron/proton (n/p) spectra in antiproton induced nuclear reactions are investigated within the LQMD transport model. In the model, the dynamics of resonances with the mass below 2 GeV, hyperons and mesons is coupled to the hadron-hadron collisions, antibaryon-baryon annihilations, decays of resonances \cite{Fe14,Fe11}. The temporal evolutions of all particles are described by Hamilton's equations of motion under the self-consistently generated mean-field potentials. The Hamiltonian of nucleons and nucleonic resonances is constructed from the Skyrme energy-density functional. Dynamics of hyperons, anti-baryons and mesons is described with the effective interactions based on the relativistic covariant theories.

The interaction potential of nucleonic system is evaluated from the energy-density functional of
$U_{loc}=\int V_{loc}(\rho(\mathbf{r}))d\mathbf{r}$ with
\begin{eqnarray}
V_{loc}(\rho)=&& \frac{\alpha}{2}\frac{\rho^{2}}{\rho_{0}} +
\frac{\beta}{1+\gamma}\frac{\rho^{1+\gamma}}{\rho_{0}^{\gamma}} + E_{sym}^{loc}(\rho)\rho\delta^{2}
\nonumber \\
&& + \frac{g_{sur}}{2\rho_{0}}(\nabla\rho)^{2} + \frac{g_{sur}^{iso}}{2\rho_{0}}[\nabla(\rho_{n}-\rho_{p})]^{2},
\end{eqnarray}
where the $\rho_{n}$, $\rho_{p}$ and $\rho=\rho_{n}+\rho_{p}$ are the neutron, proton and total densities, respectively, and the $\delta=(\rho_{n}-\rho_{p})/(\rho_{n}+\rho_{p})$ being the isospin asymmetry of baryonic matter. The parameters $\alpha$, $\beta$ and $\gamma$ are taken to be -226.5 MeV, 173.7 MeV and 1.309, respectively. The set  of the parameters gives the compression modulus of K=230 MeV for isospin symmetric nuclear matter at the saturation density ($\rho_{0}=0.16$ fm$^{-3}$). The surface coefficients $g_{sur}$ and $g_{sur}^{iso}$ are taken to be 23 MeV fm$^{2}$ and -2.7 MeV fm$^{2}$, respectively. The third term contributes the symmetry energy being of the form $E_{sym}^{loc}=\frac{1}{2}C_{sym}(\rho/\rho_{0})^{\gamma_{s}}$. The parameter $C_{sym}$ is taken as the value of 38 MeV. The $\gamma_{s}$ could be adjusted to get the suitable case from constraining the isospin observables, e.g., the values of 0.5, 1 and 2 being the soft, linear and hard symmetry energy, respectively. Combined the kinetic energy from the isospin difference of nucleonic Fermi motion, the three kinds cross at the saturation density with the value of 31.5 MeV. The mean-field potential of antinucleon is constructed from the G-parity transformation of nucleon self-energies with a scaling approach \cite{Fe14,Bu12}, which leads to the strength of optical potential $V_{\overline{N}}=-164$ MeV at the normal nuclear density.

The annihilation reactions in antibaryon-baryon collisions are described by a statistical model with SU(3) symmetry of pseudoscalar and vector mesons \cite{Go92}, which considers possible combinations with the final state from two to six mesons \cite{La12}. Pions as the dominant products in antibaryon-baryon annihilation contribute the energy deposition into the target nucleus via the pion-nucleon collisions, which leads to the emissions of pre-equilibrium particles, fission, evaporation of nucleons and light fragments etc. The cross section of pion-nucleon scattering is evaluated with the Breit-Wigner formula as the form of \cite{Li01}
\begin{eqnarray}
\sigma_{\pi N\rightarrow R}(\sqrt{s}) = \sigma_{max}|\textbf{p}_{0}/\textbf{p}|^{2}\frac{0.25\Gamma^{2}(\textbf{p})}
{0.25\Gamma^{2}(\textbf{p})+(\sqrt{s}-m_{0})^{2}},
\end{eqnarray}
where the $\textbf{p}$ and $\textbf{p}_{0}$ are the momenta of pions at the energies of $\sqrt{s}$ and $m_{0}$, respectively, and $m_{0}$ being the centroid of resonance mass, e.g., 1.232 GeV for $\Delta$(1232). The maximum cross section $\sigma_{max}$ is taken from fitting the available experimental data and satisfied to the square of Clebsch-Gordan coefficients \cite{Fe16a}, i.e., 200 mb, 133.3 mb, and 66.7 mb for $\pi^{+}+p\rightarrow \Delta^{++}$, $\pi^{0}+p\rightarrow \Delta^{+}$ and $\pi^{-}+p\rightarrow \Delta^{0}$, respectively.

\begin{figure*}
\includegraphics[width=16 cm]{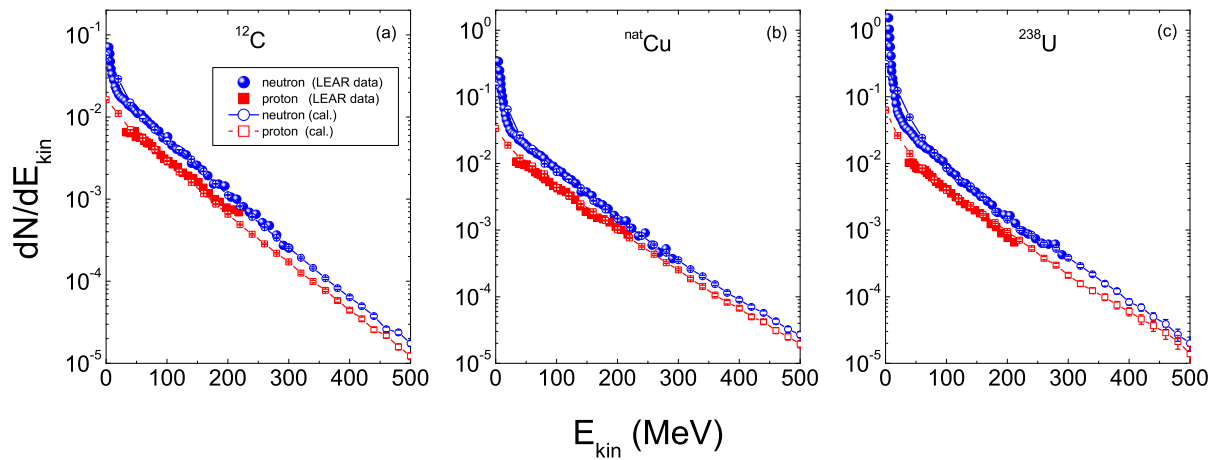}
\caption{(color online). Kinetic energy spectra of neutrons and protons produced in antiproton annihilations on carbon, copper and uranium at the incident momentum of 200 MeV/c and compared with the available data at the LEAR facility \cite{Po95}.}
\end{figure*}

The emission of fast nucleons produced in antiproton induced reactions is significant observable in understanding the energy deposition, antiproton-nucleon and meson-nucleon interactions in nuclear medium. Shown in Fig. 1 is the kinetic energy distributions of neutrons and protons produced in antiproton annihilations on $^{12}$C, $^{nat}$Cu and $^{238}$U at the incident momentum of 200 MeV/c and compared with the available data at the LEAR facility \cite{Po95}. Overall, the spectra can be nicely understood within the LQMD transport model. The free nucleons in the model are constructed with a coalescence approach, in which nucleons at the freeze-out stage in phase space are considered to belong to one cluster with the relative momentum smaller than $P_{0}$ and with the relative distance smaller than $R_{0}$ (here $P_{0}$ = 200 MeV/c and $R_{0}$ = 3 fm). The nucleon yields are weakly influenced by varying the coalescence parameters. The antiproton-nucleus systems evolve to 500 fm/c for judging the free nucleon formation. It should be noticed that the neutrons are preferable to be emitted in comparison to protons. The energy deposition in antiproton induced reactions is more explosive than the case in Fermi-energy heavy-ion collisions \cite{Co16,Fe16b}, which leads to the energetic nucleon emission. In the annihilation of an antiproton in a nucleus, pions are the dominant products. For example, the multiplicities of $\pi^{-}$ and $\pi^{+}$ on the target of $^{12}$C are 1.5 and 0.6, respectively. On the other hand, the larger elastic scattering cross sections in the $\pi^{-}n$ reactions (the maximal value being 200 mb at the $\Delta$-resonance energy (E=0.19 GeV, p=0.298 GeV/c)) in comparison to the $\pi^{-}p$ collisions (25 mb at the pion energy of 0.19 GeV), enhance the $\pi^{-}n$ collision probabilities and are favorable to neutron emissions in the low-energy antiproton induced reactions even on the isospin symmetric target of $^{12}$C. For the neutron-rich target such as $^{238}$U, the difference of neutron and proton yields is more pronounced.

\begin{figure}
\includegraphics[width=8 cm]{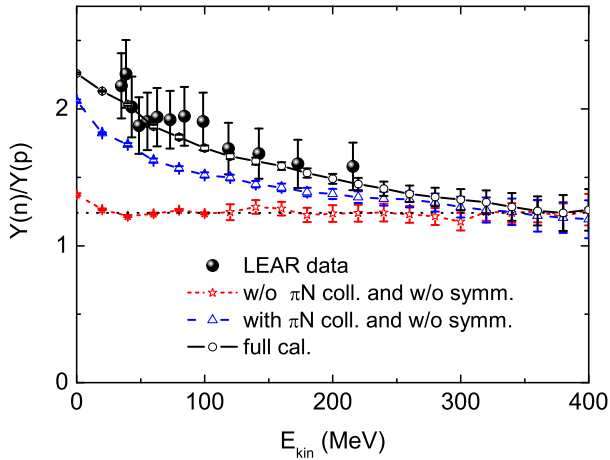}
\caption{(color online). Impacts of $\pi$N absorption and symmetry energy on the yield ratios of neutrons to protons in the $\overline{p}+^{65}$Cu reaction at 200 MeV/c and the experimental data with the stopped $\overline{p}$ on $^{\texttt{nat}}$Cu \cite{Po95}.}
\end{figure}

The ratio spectra of the isospin particles produced in heavy-ion collisions are related to the isospin dependent cross sections and interaction potentials \cite{Fe11,Ba05,Li08}. The kinetic energy distributions of the n/p ratio and double n/p ratio in isotopic nuclear reactions have been used for constraining the density dependence of symmetry energy. The yield ratios of neutrons and protons are complicated in the low-energy antiproton induced reactions, which are related to the antiproton-nucleon scattering, annihilation in antinucleon-nucleon collisions and pion-nucleon scattering. Although the experimental n/p spectra in pion- and proton-nucleus collisions were nicely explained by the intranuclear cascade model \cite{Ps95}. The kinetic energy spectra of the n/p ratio in antiproton induced reactions can not be understood by the model. The n/p ratio in antiproton-nucleus collisions is further investigated within the LQMD transport model, in which the isospin effects associated with the pion and nucleon dynamics are treated self-consistently. The yield ratios of neutrons and protons produced in the $\overline{p}+^{65}$Cu collisions at the incident momentum of 200 MeV/c are strongly constrained by the pion-nucleon collisions and symmetry energy as shown in Fig. 2. Here, the linear symmetry energy with $\gamma_{s}=$1 is chosen. Pions as the dominant products in the antiproton annihilation in a nucleus, contribute the nucleon removal from the target nucleus \cite{Fe16c}. A flat n/p spectrum appears once removed the pion-nucleon collisions and is close to the mean value (1.24) of target nucleus. The $n\pi^{-}$ scattering enhances the neutron emission, in particular at the kinetic energies below 150 MeV. The symmetry energy further increases the n/p ratio because of its repulsive contribution to neutrons in the neutron-rich matter. Both the pion-nucleon collisions and the symmetry energy dominate the n/p spectra.

\begin{figure*}
\includegraphics[width=16 cm]{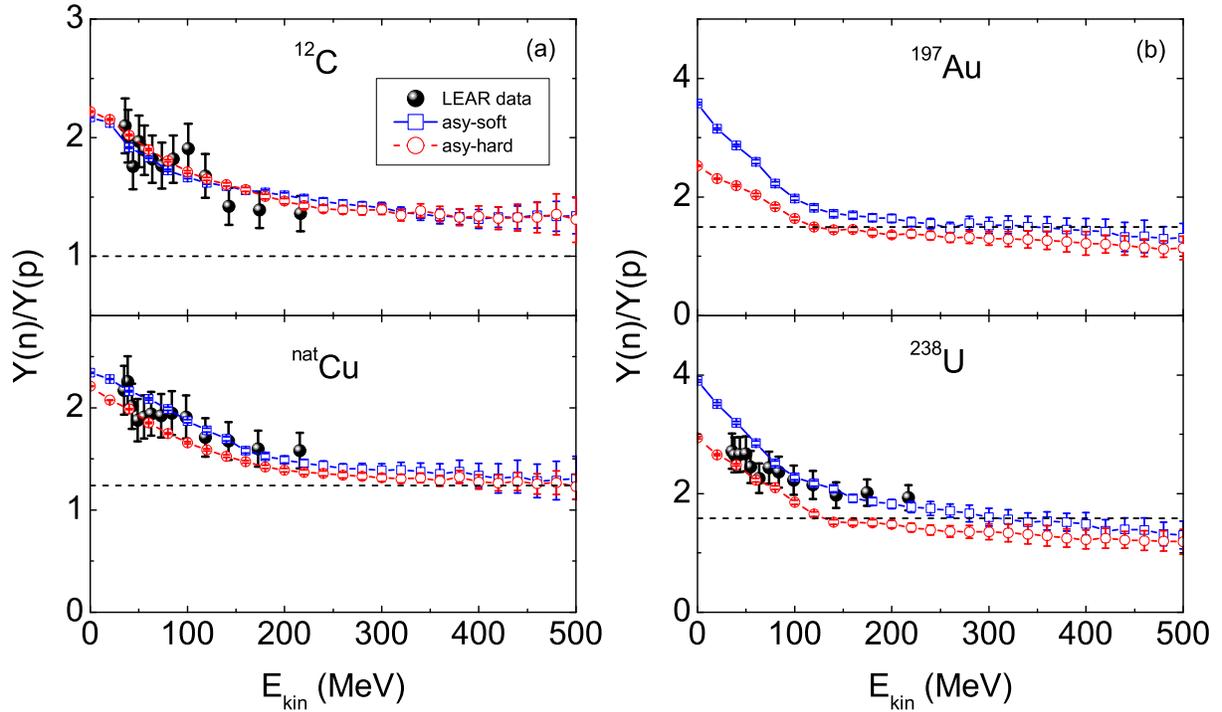}
\caption{(color online). The n/p ratios in antiproton induced reactions on $^{12}$C, $^{\texttt{nat}}$Cu, $^{197}$Au and $^{238}$U with different stiffness of symmetry energies and compared with the LEAR data \cite{Po95}.}
\end{figure*}

The stiffness of symmetry energy impacts the isospin dynamics in heavy-ion collisions. However, the nuclear density of isospin particle emissions varies in the evolution of nucleus-nucleus collisions. On the other hand, the rotation of colliding system complicates the emission angles of isospin particles. The low-energy antiproton is annihilated on the surface of the target nucleus, in which the nucleons are emitted around the density of 0.1 fm$^{-3}$ after interacting with pions. The n/p ratio directly provides the symmetry energy information at subsaturation densities. Shown in Fig. 3 is the kinetic energy spectra in antiproton induced reactions on $^{12}$C, $^{\texttt{nat}}$Cu, $^{197}$Au and $^{238}$U with the soft ($\gamma_{s}=$0.5) and hard ($\gamma_{s}=$2) symmetry energies, respectively. The dashed lines represent the mean n/p values of target nuclei. Calculations are performed at the antiproton momentum of 200 MeV/c. I have checked that the spectra are insensitive to the incident energy. It is obvious that the difference of the stiffness of symmetry energy is pronounced in the neutron-rich nuclei. The n/p ratio is enhanced with softening the symmetry energy. Overall, the available data at LEAR \cite{Po95} are reproduced with the soft symmetry energy. Different with the Fermi-energy heavy-ion collisions \cite{Fe16b}, the isospin effect appears at the kinetic energies below 200 MeV. The kinetic energy spectra of the n/p ratio in the antiproton induced reactions is expected to be further measured at PANDA (Antiproton Annihilation at Darmstadt, Germany) in the near future experiments.

In conclusion, the kinetic energy spectra of the n/p ratio produced in the antiproton annihilation in the nucleus have been puzzling for several decades. The structure is quite different with the proton (pion) induced reactions and also with the heavy-ion collisions. The available data from the LEAR facility are nicely explained with the LQMD transport model for the first time. It is found that the $n\pi^{-}$ scattering and the symmetry energy increase the neutron emission and lead to the enhancement of the ratio, in particular in the domain of kinetic energies below 200 MeV because of the larger $n\pi^{-}$ collision probability. The soft symmetry energy with the stiffness of $\gamma_{s}=$0.5 is constrained from the LEAR data.

\textbf{Acknowledgements}

This work was supported by the Major State Basic Research Development Program in China (2015CB856903), and the National Natural Science Foundation of China (Nos 11675226, 11175218 and U1332207).

\end{document}